# The phase structure of pure Regge gravity *

W. Beirl[a], A. Hauke[a], P. Homolka[ab], B. Krishnan[a], H. Kröger[b], H. Markum[a] and J. Riedler[a]

[a]Institut für Kernphysik, Technische Universität Wien, A-1040 Vienna, Austria

[b]Département de Physique, Université Laval, Québec G1K 7P4, Canada

We examine the phase structure of pure Regge gravity in four dimensions and compare our Monte Carlo results with $Z_2$-link Regge-theory as well as with another formulation of lattice gravity derived from group theoretical considerations. Within all three models we find an extension of the well-defined phase to negative gravitational coupling and a new phase transition. In contrast to the well-known transition at positive coupling there is evidence for a continuous phase transition which might be essential for a possible continuum limit.

## 1. Introduction

It is the aim of this work to present an extension of the phase diagram of four dimensional simplicial quantum gravity in the Regge approach. Due to the fact that a Wick rotation to the Euclidean sector of quantum gravity is not feasible in general the sign in the exponential of the path integral is not fixed a priori. We investigate the continuation of the well-defined, smooth phase [1,2] into the region of negative gravitational coupling and its termination by a transition to an ill-defined, rough phase with very large negative expectation values of the curvature. We further confront this phase diagram with those of two more lattice gravity models being introduced below.

## 2. Lattice quantum gravity

In our numerical simulations we evaluate the following path integral

$$Z = \int D[q] e^{-I(q)} \ , \tag{1}$$

with the quadratic link lengths $q_l$ as the dynamical degrees of freedom [3]. We refer to the literature [2,4] concerning the various problems of (1). On the lattice there is a certain freedom in defining the action $I(q)$, the measure $D[q]$, and the underlying lattice structure itself. Mainly for computational simplicity we use a regularly triangulated hypercubic lattice with toroidal topology and $4^4$ vertices. For

$$D[q] = \prod_l dq_l \mathcal{F}(q) \ , \tag{2}$$

we choose the simple uniform measure. The function $\mathcal{F}$ ensures that only Euclidean link configurations are taken into account, i.e. $\mathcal{F} = 1$ if the Euclidean triangle inequalities are fulfilled and $\mathcal{F} = 0$ otherwise.

Now we define the three different versions of the action under investigation.

### 2.1. Model 1: Conventional Regge gravity

Here we employ the Regge-Einstein action inclusive a cosmological term [5]

$$I_r = -\beta \sum_t A_t \delta_t + \lambda \sum_s V_s \ . \tag{3}$$

The first sum runs over all products of triangle area $A_t$ times corresponding deficit angle $\delta_t$ weighted by the bare gravitational coupling $\beta$. The second sum extends over the volumes $V_s$ of the 4-simplices of the lattice and allows together with the cosmological constant $\lambda$ to set an overall scale in the action. In our current simulations we have fixed $\lambda = 1$.

It should be mentioned that the Regge action $2\sum_t A_t \delta_t$ is equivalent to the Einstein-Hilbert action $\int d^4x \sqrt{g} R$ in the classical continuum limit if

*Supported in part by Fonds zur Förderung der wissenschaftlichen Forschung under Contract P9522-PHY.

the so-called fatness $\phi_s$ of a 4-simplex obeys [6]

$$\phi_s \sim \frac{V_s^2}{\max_{l\in s}(q_l^4)} \geq f = \text{const} > 0 \ . \qquad (4)$$

A lower limit $f = 10^{-4}$ restricts the configuration space and thus facilitates numerical simulations [4]. We performed at least 50k Monte Carlo sweeps for each value of the coupling $\beta$.

### 2.2. Model 2: Group theoretical approach

Given a simplicial lattice, to each link of its dual lattice a Poincaré transformation can be assigned. With these new dynamical variables belonging to the Poincaré group it is possible to construct an action

$$I_c = -\beta \sum_t A_t \sin\delta_t + \lambda \sum_s V_s \qquad (5)$$

that reduces to (3) in the small curvature limit [7]. The higher order terms from an expansion of $\sin\delta_t$ do not affect the classical continuum limit [8]. Again we have used $\lambda = 1$ and $f = 10^{-4}$ in our Monte Carlo simulations with at least 50k iterations.

### 2.3. Model 3: $Z_2$-link Regge-gravity

This model was invented in an attempt to reformulate (1) as the partition function of a spin system [9,10]. It is defined by restricting the squared link lengths to take on only two values

$$q_l = b_l(1+\epsilon\sigma_l) \ , \quad \sigma_l \in Z_2 \ . \qquad (6)$$

By setting $b_l = 1,2,3,4$ for edges, face diagonals, body diagonals and the hyperbody diagonal of a hypercube, respectively, the link lengths are allowed to fluctuate around their flat-space values. The Euclidean triangle inequalities are automatically fulfilled as long as $0 \leq \epsilon < \epsilon_{max}$ and therefore $\mathcal{F} = 1$ in any case. Consequently, the measure (2) in the quantum-gravity path-integral becomes identical to one for all possible link configurations. The action (3) can be rewritten in terms of complicated local "spin" interactions but the result is not particularly illustrative and we omit the details here.

Numerical simulations of the $Z_2$ system become extremely efficient by implementing look-up tables and a heat-bath algorithm. A few thousand iterations are taken into account after reaching thermal equilibrium. Computations have been performed with the parameter $\epsilon = 0.0875$ and the cosmological constant $\lambda = 0$ because (6) already fixes the average lattice volume.

## 3. Results

A quantity of interest in lattice gravity is the average link length

$$\bar{q} = \frac{1}{N_1}\sum_l q_l \ . \qquad (7)$$

In the $Z_2$-link model it is related to an important order parameter, the magnetization $\bar{\sigma}$. Both quantities are not geometrically invariant. Another important observable is the scale invariant average curvature [2]

$$\bar{R}_r = \frac{2\sum_t A_t\delta_t}{\sum_s V_s}\bar{q} \qquad (8)$$

or

$$\bar{R}_c = \frac{2\sum_t A_t \sin\delta_t}{\sum_s V_s}\bar{q} \qquad (9)$$

of a configuration. $N_i$ denotes the total number of $i$-simplices. As $\beta$ is varied all three models undergo phase transitions at two different points.

First let us concentrate on the Regge approach (Model 1), cf. Fig. 1a. The expectation value of the average link length (7) changes only slightly but increases considerably near $\beta_c^- \approx -0.16$, $\beta_c^+ \approx 0.116$ and becomes arbitrarily large for $\beta_c^- > \beta > \beta_c^+$ indicating the emergence of spike-like structures. The expectation value of the average curvature (8) is negative in the well-defined or smooth phase, but $|\langle\bar{R}_r\rangle|$ becomes very large in the ill-defined or rough phase where simplices collapse into degenerate configurations with long links and small volumes.

Model 2 behaves very similar, see Fig. 1b. Again the expectation values of the average link length (7) and of the average curvature (9) are finite and well-defined only in a certain coupling region. An additional kink is observed at $\beta \approx -0.115$ and recent simulations seem to indicate that the points for smaller $\beta$-values become unstable for $f \to 0$.

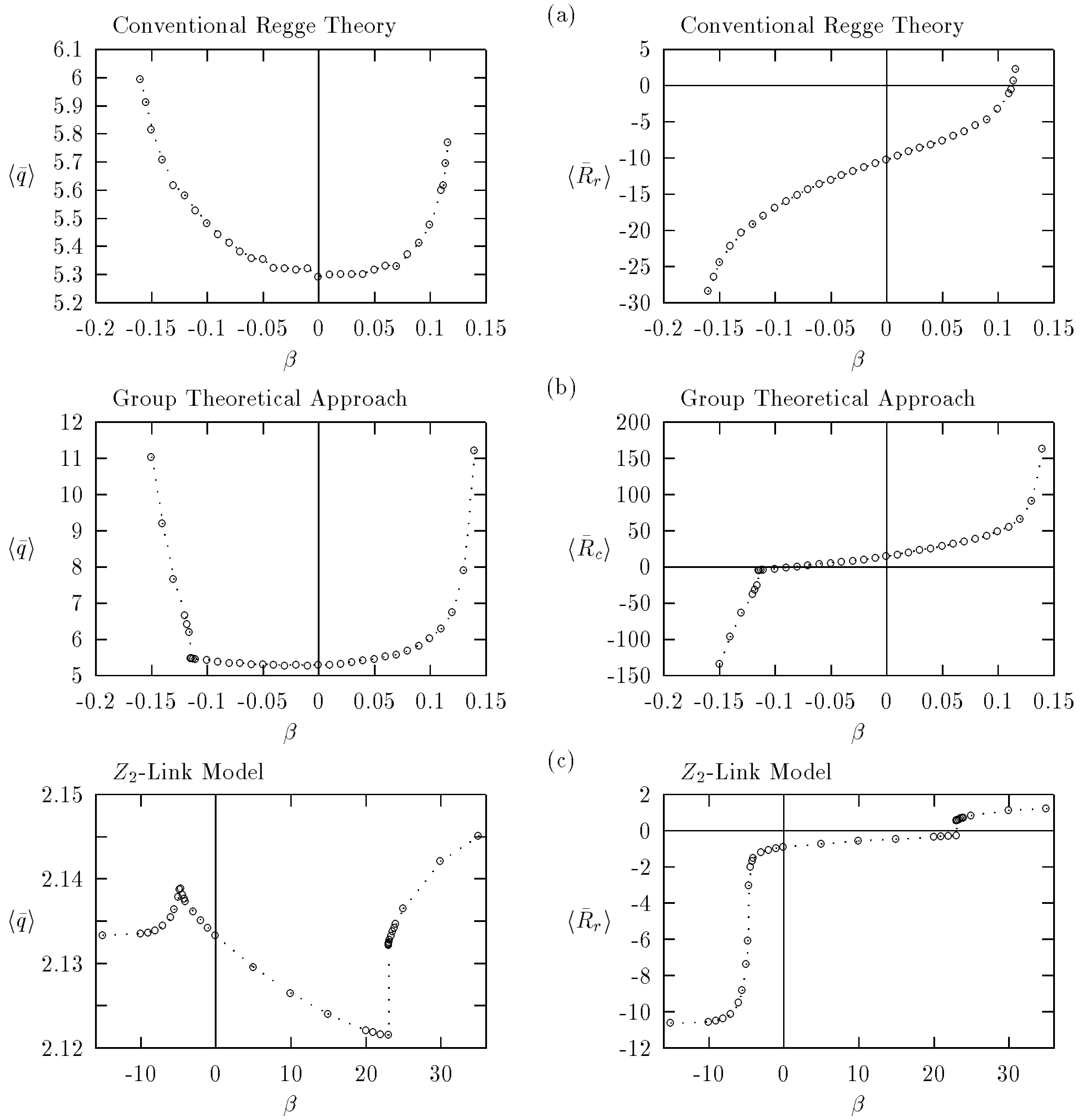

Figure 1. Expectation values of the average link lengths (left plots) and the average curvatures (right plots) as a function of the gravitational coupling for (a) the Regge theory, (b) the group theoretical approach, and (c) the $Z_2$-link model. All models exhibit a related phase structure and could lie in the same universality class. Error bars are in the size of symbols except near the transition points.

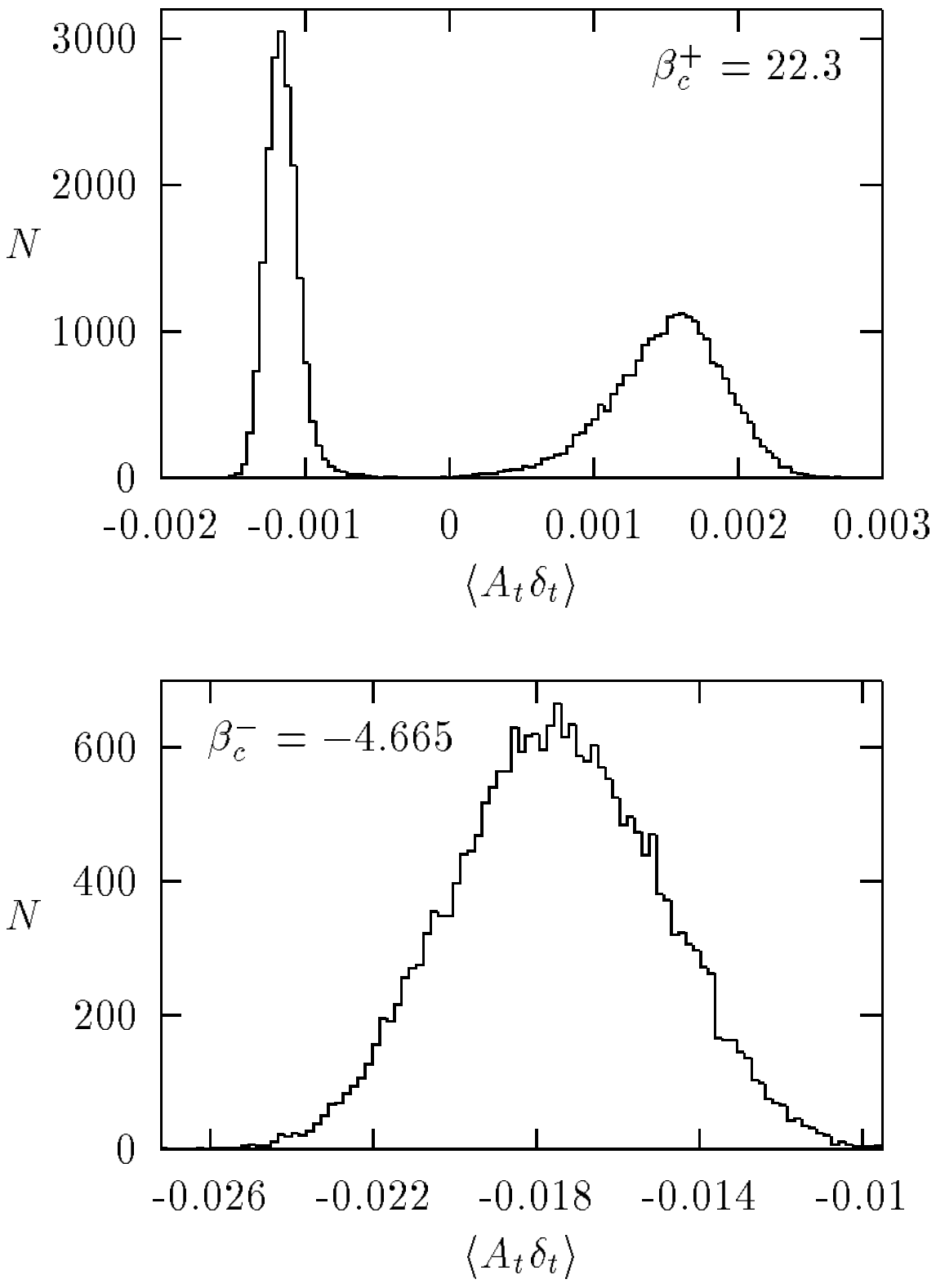

Figure 2. Histograms of the Regge action $\langle A_t \delta_t \rangle$ from simulations of $Z_2$-link Regge-gravity. The distributions indicate at $\beta_c^+ = 22.3$ a first order transition (upper plot) and at $\beta_c^- = -4.665$ a continuous transition (lower plot).

Therefore we assume the well-defined phase to be located between $\beta_c^- \approx -0.115$ and $\beta_c^+ \approx 0.14$. It is remarkable that the curvature is mostly positive in this coupling interval.

In contrast the $Z_2$-link Regge-gravity (Model 3) is always well-defined because (6) limits the link lengths. The lattice freezes forming characteristic configurations in the corresponding ill-defined phases of Model 1 and 2, cf. Fig. 1c.

Such a system is of course better suited for investigations of the phase transitions. Fig. 2 depicts histograms of the action observable $\langle A_t \delta_t \rangle$ at the critical values of the coupling. For $\beta_c^+ = 22.3$ we get a typical two-peak structure indicating a first order transition whereas for $\beta_c^- = -4.665$ there is only one peak visible hinting at a continuous phase transition. Its order is currently examined considering observables like the specific heat by means of histogram and finite size scaling techniques and applying the Binder-Challa-Landau criterium.

## 4. Summary

We explored the complete phase structure of three different formulations of lattice quantum gravity in four dimensions including the region of negative gravitational couplings. The resemblance of the results from simulations of $Z_2$-link Regge-theory (Model 3) to those obtained by continuously varying edge lengths (Model 1) is of particular interest. We found a first order phase transition at $\beta_c^+ > 0$ in the $Z_2$-model which is reported for conventional Regge quantum-gravity as well [2]. But, to show whether both models belong to the same universality class critical exponents at $\beta_c^- < 0$ have to be extracted carefully.